\documentclass[twocolumn]{article}

\usepackage[utf8]{inputenc}
\usepackage[T1]{fontenc}
\usepackage{lmodern}
\usepackage{graphicx}
\usepackage{amsmath,amssymb}
\usepackage{booktabs}
\usepackage[hyphens]{url}
\usepackage[hidelinks,breaklinks=true]{hyperref}
\usepackage{authblk} 

\title{Adaptive Learning Strategies for Mitotic Figure Classification in MIDOG2025 Challenge}

\author[1]{Biwen Meng}
\author[1]{Xi Long}
\author[1]{Jingxin Liu}
\affil[1]{School of AI and Advanced Computing, Xi'an Jiaotong-Liverpool University}
\date{} 

\begin{document}
\maketitle

\begin{abstract}
Atypical mitotic figures (AMFs) are clinically relevant indicators of abnormal cell division, yet their reliable detection remains challenging due to morphological ambiguity and scanner variability. In this work, we investigated three variants of adapting the pathology foundation model UNI2 for the MIDOG2025 Track 2 challenge: (1) LoRA + UNI2, (2) VPT + UNI2 + Vahadane Normalizer, and (3) VPT + UNI2 + GRL + Stain TTA. We observed that the integration of Visual Prompt Tuning (VPT) with stain normalization techniques contributed to improved generalization. The best robustness was achieved by further incorporating test-time augmentation (TTA) with Vahadane and Macenko stain normalization. Our final submission achieved a balanced accuracy of 0.8837 and an ROC-AUC of 0.9513 on the preliminary leaderboard, ranking within the top 10 teams. These results suggest that prompt-based adaptation combined with stain-normalization TTA offers a promising strategy for atypical mitosis classification under diverse imaging conditions.
\end{abstract}


\section*{Introduction}
The density of mitotic figures (MFs) in histopathological tumor specimens is highly correlated with tumor proliferation and is regarded as an important criterion for tumor grading \cite{aubreville2023mitosis, veta2019predicting}. In routine H\&E-stained slides, mitotic cells can be detected and counted within specific tumor regions. However, not all mitotic figures are morphologically similar. Atypical mitotic figures (AMFs) represent cells undergoing abnormal division, often characterized by chromosome segregation errors or irregular morphological features \cite{banerjee2025benchmarking}. Distinguishing AMFs from typical mitoses or apoptotic cells remains highly challenging due to their subtle and heterogeneous appearance. Recent studies have highlighted the clinical relevance of AMFs. For example, Jahanifar et al. (2025) \cite{jahanifar2025pan} provided large-scale evidence that the frequency of AMFs across different tumor types is associated with patient prognosis, underscoring the importance of investigating this parameter further. Nevertheless, manual identification of AMFs is time-consuming, prone to inter-observer variability, and limited in scalability. The inter-rater agreement for classifying MFs into normal and atypical subtypes has been reported as low, reflecting the complexity and variability of their morphologies\cite{bertram2025histologic}.

Deep learning approaches have shown promise for atypical mitotic figure classification by improving reproducibility and reducing the time investment. Nevertheless, their performance is often hindered by the challenges mentioned above: the scarcity of mitotic figures amplifies class imbalance, the complex and heterogeneous morphologies increase intra-class variability, and the subtle distinction between normal and atypical subtypes complicates reliable classification. The MIDOG2025 challenge track 2 provides a benchmark specifically designed to evaluate robustness under these conditions. By focusing on the classification of atypical mitotic figures in diverse test scenarios, it enables systematic comparison of algorithms and drives the development of more reliable methods for computational pathology.

In this work, we propose a framework for atypical mitosis classification that leverages the pathology foundation model UNI2-h with lightweight prompt tuning to capture discriminative morphological features. To address variability introduced by different scanners, adversarial training with weak domain labels is applied to encourage the learning of domain-invariant representations. Additionally, stain normalization is incorporated to reduce appearance discrepancies across samples. The proposed method was systematically evaluated in the MIDOG2025 Track 2 challenge, where it achieved strong performance on both the validation and preliminary test sets, ranking within the top 10 on the official leaderboard.

\section*{Methods}
\begin{figure*}[t]
  \centering
  \includegraphics[width=0.95\textwidth]{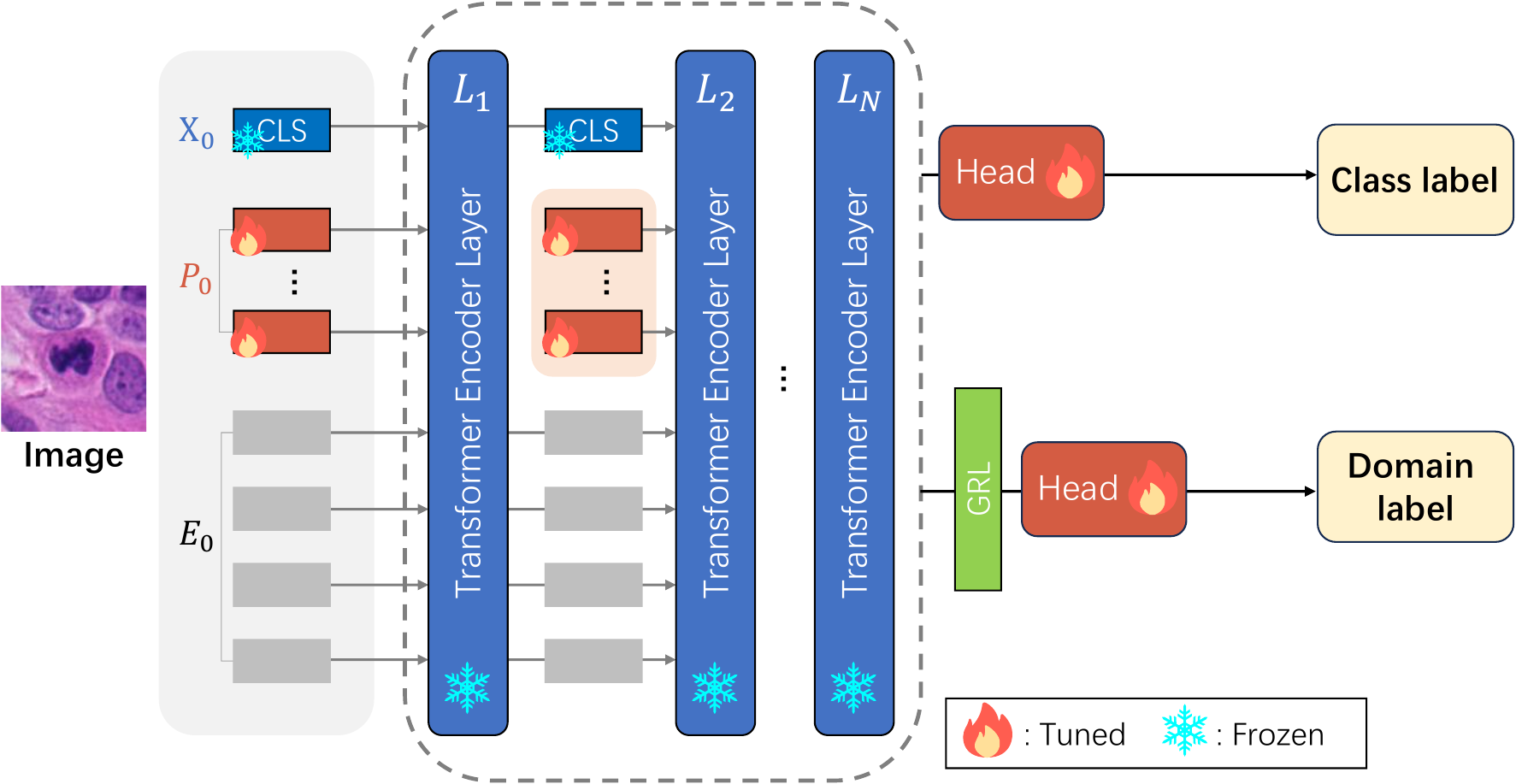}
  \caption{Overview of the proposed framework for atypical mitosis classification. The input H\&E image is first mapped by the patch embedding layer to obtain patch-level representations $E_0$. A learnable class token $X_0$ is added to aggregate global semantics, and learnable prompt tokens $P_0$ are inserted before each transformer encoder layer to modulate feature extraction. A frozen UNI2-h backbone with trainable prompt tokens processes the input through transformer encoder layers. The shared features are passed to a classification head for predicting mitotic subtype labels and to a domain-adversarial branch consisting of a Gradient Reversal Layer (GRL) and fully connected layers for scanner domain prediction.}
  \label{fig:framework}
\end{figure*}
Our proposed framework for atypical mitosis classification is illustrated in Figure \ref{fig:framework}. It is built upon the pathology foundation model UNI2-h\cite{chen2024towards}, which provides strong histopathological representations through a transformer-based backbone. To adapt the model for atypical mitosis classification under scanner variability, we introduce three complementary strategies: (1) visual prompt tuning (VPT) for efficient adaptation, (2) adversarial domain alignment to reduce scanner-specific biases, and (3) stain normalization and augmentation for appearance-level robustness.

\subsection{Visual Prompt Tuning on UNI2-h}
We adopt Visual Prompt Tuning (VPT) \cite{jia2022visual} on UNI2-h. In this design, Class tokens ($X_0$) serve as learnable global classification tokens, while prompt tokens ($P_0$) are inserted before each transformer encoder block ($L_1, …, L_N$) and removed after processing. The patch embeddings ($E_0$) retain local morphological information. During training, the backbone parameters are frozen, while only the prompt tokens and the classification head are updated. This greatly reduces trainable parameters while still enabling the model to capture discriminative morphological patterns of atypical mitoses.

\subsection{Domain-Adversarial Learning}
To explicitly encourage domain-invariant features, we attach a domain classifier to the shared feature space and train it with a Gradient Reversal Layer (GRL) \cite{ganin2015unsupervised}. Scanner labels are used to supervise this branch, while the adversarial loss penalizes the backbone if scanner-specific information is preserved. This strategy has been widely used in domain adaptation tasks and here improves generalization to unseen scanners\cite{tzeng2017adversarial}.

\subsection{Implementation Details}
During inference, we adopt a test-time augmentation (TTA) strategy to improve the robustness of predictions. Each test patch is evaluated under multiple transformations, including horizontal and vertical flips as well as $90^\circ$ rotations. In addition to color augmentations, we apply two complementary stain normalization pipelines, namely Vahadane \cite{vahadane2016structure} and Macenko \cite{macenko2009method}. For each input, predictions across all augmented versions are averaged to obtain the final probability. This ensemble-like strategy reduces prediction variance and improves generalization across scanners.

\section*{Results}

\begin{table}[t]
\centering
\caption{Track 2 (Atypical Mitosis Classification) Preliminary leaderboard results.}
\label{tab:leaderboard}
\resizebox{0.9\linewidth}{!}{%
\begin{tabular}{lcccc}
\toprule
Method & Balanced Acc. & Sensitivity & Specificity & ROC AUC \\
\midrule
LoRA + UNI2-h & 0.8305 & 0.8169 & 0.8443 & 0.9364 \\
VPT + UNI2-h + VahadaneNorm & 0.8711 & 0.9014 & 0.8408 & 0.9483 \\
\textbf{VPT + UNI2-h + GRL + Stain TTA} & \textbf{0.8837} & \textbf{0.9577} & \textbf{0.8097} & \textbf{0.9513} \\
\bottomrule
\end{tabular}
}
\end{table}
We compared three variants of our framework on the MI-DOG2025 Track 2 dataset. As shown in Table \ref{tab:leaderboard}, the LoRA + UNI2-h baseline achieved a balanced accuracy of 0.8305 and an ROC-AUC of 0.9364. Replacing LoRA with Visual Prompt Tuning (VPT) along with Vahadane stain normalization improved the results, yielding a balanced accuracy of 0.8711 and an ROC-AUC of 0.9483. Finally, incorporating a Gradient Reversal Layer (GRL) for domain adaptation and test-time augmentation (TTA) with both Vahadane and Macenko normalization further enhanced robustness, achieving a balanced accuracy of 0.8837 and the highest ROC-AUC of 0.9513. Based on these results, we selected the variant combining VPT, GRL, and TTA with UNI2-h as our final submission.


\section*{Conclusion}
In this study, we investigated three variants of adapting the UNI2-h foundation model for atypical mitosis classification in the MIDOG2025 Track 2 challenge. The LoRA-based approach or VPT provided a competitive baseline, while stain normalization improved generalization. Further integrating GRL along with test-time augmentation yielded the best results, achieving a balanced accuracy of 0.8837 and an ROC-AUC of 0.9513 on the preliminary test set. These findings demonstrate that combining prompt-based adaptation, stain normalization, and adversarial domain adaptation is an effective strategy for robust mitosis classification under diverse imaging conditions.


\bibliographystyle{elsarticle-num}
\bibliography{literature}

\begin{thebibliography}{10}
\expandafter\ifx\csname url\endcsname\relax
  \def\url#1{\texttt{#1}}\fi
\expandafter\ifx\csname urlprefix\endcsname\relax\def\urlprefix{URL }\fi
\expandafter\ifx\csname href\endcsname\relax
  \def\href#1#2{#2} \def\path#1{#1}\fi

\bibitem{aubreville2023mitosis}
M.~Aubreville, N.~Stathonikos, C.~A. Bertram, R.~Klopfleisch, N.~Ter~Hoeve, F.~Ciompi, F.~Wilm, C.~Marzahl, T.~A. Donovan, A.~Maier, et~al., Mitosis domain generalization in histopathology images—the midog challenge, Medical Image Analysis 84 (2023) 102699.

\bibitem{veta2019predicting}
M.~Veta, Y.~J. Heng, N.~Stathonikos, B.~E. Bejnordi, F.~Beca, T.~Wollmann, K.~Rohr, M.~A. Shah, D.~Wang, M.~Rousson, et~al., Predicting breast tumor proliferation from whole-slide images: the tupac16 challenge, Medical image analysis 54 (2019) 111--121.

\bibitem{banerjee2025benchmarking}
S.~Banerjee, V.~Weiss, T.~A. Donovan, R.~H. Fick, T.~Conrad, J.~Ammeling, N.~Porsche, R.~Klopfleisch, C.~Kaltenecker, K.~Breininger, et~al., Benchmarking deep learning and vision foundation models for atypical vs. normal mitosis classification with cross-dataset evaluation, arXiv preprint arXiv:2506.21444 (2025).

\bibitem{jahanifar2025pan}
M.~Jahanifar, M.~Dawood, N.~Zamanitajeddin, A.~Shephard, B.~S. Chohan, C.~A. Bertram, N.~Wahab, M.~Eastwood, M.~Aubreville, S.~E.~A. Raza, et~al., Pan-cancer profiling of mitotic topology \& mitotic errors: Insights into prognosis, genomic alterations, and immune landscape, medRxiv (2025) 2025--06.

\bibitem{bertram2025histologic}
C.~A. Bertram, V.~Weiss, T.~A. Donovan, S.~Banerjee, T.~Conrad, J.~Ammeling, R.~Klopfleisch, C.~Kaltenecker, M.~Aubreville, Histologic dataset of normal and atypical mitotic figures on human breast cancer (ami-br), in: BVM Workshop, Springer, 2025, pp. 113--118.

\bibitem{chen2024towards}
R.~J. Chen, T.~Ding, M.~Y. Lu, D.~F. Williamson, G.~Jaume, A.~H. Song, B.~Chen, A.~Zhang, D.~Shao, M.~Shaban, et~al., Towards a general-purpose foundation model for computational pathology, Nature medicine 30~(3) (2024) 850--862.

\bibitem{jia2022visual}
M.~Jia, L.~Tang, B.-C. Chen, C.~Cardie, S.~Belongie, B.~Hariharan, S.-N. Lim, Visual prompt tuning, in: European conference on computer vision, Springer, 2022, pp. 709--727.

\bibitem{ganin2015unsupervised}
Y.~Ganin, V.~Lempitsky, Unsupervised domain adaptation by backpropagation, in: International conference on machine learning, PMLR, 2015, pp. 1180--1189.

\bibitem{tzeng2017adversarial}
E.~Tzeng, J.~Hoffman, K.~Saenko, T.~Darrell, Adversarial discriminative domain adaptation, in: Proceedings of the IEEE conference on computer vision and pattern recognition, 2017, pp. 7167--7176.

\bibitem{vahadane2016structure}
A.~Vahadane, T.~Peng, A.~Sethi, S.~Albarqouni, L.~Wang, M.~Baust, K.~Steiger, A.~M. Schlitter, I.~Esposito, N.~Navab, Structure-preserving color normalization and sparse stain separation for histological images, IEEE transactions on medical imaging 35~(8) (2016) 1962--1971.

\bibitem{macenko2009method}
M.~Macenko, M.~Niethammer, J.~S. Marron, D.~Borland, J.~T. Woosley, X.~Guan, C.~Schmitt, N.~E. Thomas, A method for normalizing histology slides for quantitative analysis, in: 2009 IEEE international symposium on biomedical imaging: from nano to macro, IEEE, 2009, pp. 1107--1110.

\end{thebibliography}

\end{document}